\documentclass{aa}
\usepackage{psfig}

\begin{document}

\title{Unveiling the nature of {\it INTEGRAL} objects through optical 
spectroscopy. II. The nature of four unidentified sources\thanks{Based 
on observations collected at the Astronomical Observatory of Bologna in 
Loiano, Italy.}}

\author{
N. Masetti\inst{1},
E. Mason\inst{2},
L. Bassani\inst{1},
A.J. Bird\inst{3},
E. Maiorano\inst{1,4},
A. Malizia\inst{1},
E. Palazzi\inst{1},
J.B. Stephen\inst{1},
A. Bazzano\inst{5}, 
A.J. Dean\inst{3},
P. Ubertini\inst{5} and
R. Walter\inst{6}
}

\institute{
INAF -- Istituto di Astrofisica Spaziale e Fisica Cosmica di 
Bologna, Via Gobetti 101, I-40129 Bologna, Italy (formerly IASF/CNR,
Bologna)
\and
European Southern Observatory, Casilla 19001, Santiago 19, Chile
\and
School of Physics \& Astronomy, University of Southampton, Southampton, 
Hampshire, SO17 1BJ, United Kingdom  
\and
Dipartimento di Astronomia, Universit\`a di Bologna, Via Ranzani 1, 
I-40126 Bologna, Italy
\and
INAF -- Istituto di Astrofisica Spaziale e Fisica Cosmica di
Roma, Via Fosso del Cavaliere 100, I-00133 Rome, Italy (formerly 
IASF/CNR, Rome)
\and
INTEGRAL Science Data Centre, Chemin d'Ecogia 16, CH-1290 Versoix,
Switzerland
}

\offprints{N. Masetti (\texttt{masetti@bo.iasf.cnr.it)}}
\date{Received 22 August 2005; accepted 7 Novembver 2005}

\abstract{
We present the results of our optical spectrophotometric
campaign ongoing at the Astronomical Observatory of Bologna in Loiano
(Italy) on hard X-ray sources detected by {\it INTEGRAL}. We
observed spectroscopically the putative optical counterparts of four 
more {\it INTEGRAL} sources, IGR J12391$-$1610, IGR J18406$-$0539, 2E
1853.7+1534 and IGR J19473+4452. These data have allowed us to 
determine their nature, finding that IGR J12391$-$1610 (=LEDA 170194) 
and IGR J19473+4452 are Seyfert 2 galaxies at redshifts $z$ = 0.036 
and $z$ = 0.053, respectively, IGR J18406$-$0539 
(=SS 406) is a Be massive X-ray binary located at $\sim$1.1 kpc 
from Earth, and 2E 1853.7+1534 is a Type 1 Seyfert galaxy with $z$ = 
0.084. Physical parameters
for these objects are also evaluated by collecting and discussing the
available multiwavelength information. The determination of the
extragalactic nature of a substantial fraction of sources inside the 
{\it INTEGRAL} surveys underlines the importance of hard X-ray
observations for the study of background Active Galactic Nuclei located
beyond the `Zone of Avoidance' of the Galactic Plane.
}

\titlerunning{The nature of four more {\it INTEGRAL} sources}
\authorrunning{N. Masetti et al.}

\maketitle

\section{Introduction}

One of the objectives of using satellites observing in the hard X-ray band 
(above 20 keV) is to obtain all-sky maps of celestial high-energy 
emission. This allows information on the sky distribution and 
characteristics of X-ray objects to be obtained, and opens an 
observational window on new populations of sources. Previously, several 
surveys have been
performed by various spacecraft, such as {\it HEAO-1} (13--180 keV; Levine 
et al. 1984), SIGMA onboard {\it Granat} (40--800 keV; Vargas et al. 1996) 
and BATSE onboard {\it Compton-GRO} (25--160 keV; Shaw et al. 2004). These 
surveys were mostly devoted to all-sky scannings, with particular 
attention to the Galactic Plane and to the Galactic Centre. However, the 
main drawbacks of these past hard X-ray surveys were the poor
positional accuracy afforded by the available technology (typical error 
boxes were of the order of some degrees) and/or the low survey sensitivity 
($\ga$30 mCrab).

In this sense, {\it INTEGRAL} (Winkler et al. 2003) produced a 
breakthrough in all-sky mapping of hard X-ray sources in terms of 
both sensitivity and positional accuracy. Indeed, thanks to the 
capabilities of the IBIS instrument (Ubertini et al. 2003), {\it INTEGRAL} 
is able to detect hard X-ray sources at the mCrab level with a typical 
localization accuracy of 2--3$'$ (Gros et al. 2003). 
This has made it possible, for the first 
time, to resolve crowded regions such as the Galactic Centre and the 
spiral arms, and discover many new hard X-ray extragalactic objects 
beyond the Galactic Plane (the so-called `Zone of Avoidance'), where the 
massive presence of neutral hydrogen hampers observations in soft X-rays.

Since the launch of {\it INTEGRAL}, the ISGRI detector of IBIS has 
detected about 150 sources above 20 keV (Bird et al. 2004; Bassani et al. 
2004; Molkov et al. 2004; Revnivstev et al. 2004a; Krivonos et al. 
2005; Revnivstev et al. 2005; Sazonov et al. 2005) down to mCrab 
sensitivities. In the widest and 
deepest of these surveys (i.e., that of Bird et al. 2004), most of the 
detected sources match already known Galactic Low-Mass and High-Mass 
X-ray Binaries (LMXBs and HMXBs; $\sim$60\%), background Active Galactic 
Nuclei (AGNs; $\sim$4\%) and Cataclysmic Variables (CVs; $\sim$4\%). The 
remaining objects (about 23$\%$ of the sample) had no obvious 
counterpart at other wavelengths and therefore could not immediately 
be associated with any known class of high-energy emitting objects.

The majority of these unidentified sources are believed to be Galactic 
X-ray binary systems, although a few of them have turned out to be AGNs 
(e.g., Masetti et al. 2004, hereafter Paper I; Combi et al. 2005). 
However, since all these objects are hard X-ray selected and poorly known 
at other wavebands, there are serious possibilities that we might also be 
dealing with known types of sources but in peculiar evolutionary 
stages (e.g., Filliatre \& Chaty 2004; Dean et al. 2005).

In order to reduce the {\it INTEGRAL} error box, correlations with 
catalogs at longer wavelengths (soft X-ray, optical, near- and 
far-infrared, and/or radio) are needed. Indeed, cross-correlation of the 
IBIS 20--100 keV catalogue with the {\it ROSAT} database (Voges et al. 
1999) indicates a high degree of association (Stephen et al. 
2005). Moreover, it increases the positional accuracy to few arcsecs, thus 
making the optical searches much easier. Similarly, the presence of a 
radio object within the IBIS error box can again be seen as an indication 
of an association between the radio emitter and the {\it INTEGRAL} source 
(e.g., Combi et al. 2005; Paper I). However, whereas the cross-correlation 
with catalogues at other wavebands is critical in pinpointing the putative 
optical candidates, only accurate optical spectroscopy can reveal the real 
nature of the X-ray emitting object.

For this reason, we started a program to perform optical spectroscopy of 
currently unidentified IBIS/{\it INTEGRAL} sources. Among these sources we 
have selected a sample of objects for which likely candidates could be 
pinpointed. In particular, we have selected our targets on the basis of
their association with sources at other wavebands, mainly in the soft
X-ray and radio. We have already been successful in providing the optical
spectroscopic identification for 3 objects (Paper I).

Here we report results obtained
at the Astronomical Observatory of Bologna in Loiano on a further group of
four sources extracted from the forthcoming 2$^{\rm nd}$ IBIS/{\it
INTEGRAL} survey (Bassani et al. 2005; Bird et al. 2005), from the
{\it INTEGRAL} Sagittarius Arm Tangent survey (Molkov et al. 2004),
and from the {\it INTEGRAL}/{\it Chandra} minisurvey of Sazonov et al.
(2005).
In Sect. 2 we present the sample of objects selected for the observational 
campaign shown here, whereas in Sect. 3 a description of the 
observations is given; Sect. 4 reports the results for each source and a 
discusses them. Conclusions are drawn in Sect. 5.
In the following, when not explicitly stated otherwise, for our X-ray 
flux estimates we will assume a Crab-like spectrum, whereas for the 
{\it INTEGRAL} error box size a conservative 90\% confidence level radius 
of 3$'$ will be considered.

\section{The selected sample}

{\it IGR J12391$-$1610}: listed in the 2$^{\rm nd}$ IBIS survey 
(Bassani et al. 2005; Bird et al. 2005), as well as in the 
{\it INTEGRAL}/{\it Chandra} minisurvey of Sazonov et al. (2005), 
this object was detected 
by ISGRI at coordinates RA = 12$^{\rm h}$ 39$^{\rm m}$ 11$\fs$0, 
Dec = $-$16$^{\circ}$ 10$'$ 55$''$ (J2000), with fluxes
(2.0$\pm$0.4)$\times$10$^{-11}$ erg cm$^{-2}$ s$^{-1}$ and 
(5.2$\pm$0.8)$\times$10$^{-11}$ erg cm$^{-2}$ s$^{-1}$
in the 20--40 and 40--100 keV bands, respectively. Within the {\it 
INTEGRAL} error box a soft X-ray source was detected by {\it Chandra} 
at a 0.5--8 keV flux of (2.0$\pm$0.3)$\times$10$^{-12}$ erg cm$^{-2}$ 
s$^{-1}$; for details, see Sazonov et al. (2005) and Halpern (2005).
We remark that these authors labeled the source as IGR J12391$-$1612.
At the {\it Chandra} subarcsecond X-ray position (RA = 12$^{\rm h}$ 
39$^{\rm m}$ 06$\fs$29, Dec = $-$16$^{\circ}$ 10$'$ 47$\farcs$1; equinox 
J2000; error radius: $\sim$0$\farcs$6), the apparently `normal' and 
optically fairly bright ($B \sim$ 15 mag) S0-type galaxy LEDA 170194 
(Paturel et al. 2003) is present (Fig. 1, top left panel). Although a 
redshift ($z$ = 0.0367$\pm$0.0001; da Costa et al. 1998) is reported, no 
classification is available in the literature for this galaxy. However, 
its characteristics,
such as the detection at longer wavebands including near- (2MASX 
J12390630$-$1610472) and far-infrared (IRAS 12365$-$1554), 
as well as in the radio (NVSS J123906$-$161046: 39.4$\pm$1.6 
mJy at 1.4 GHz; Condon et al. 1998), suggest that it might be an active
galaxy. Its position well above the Galactic Plane ($b$ = +46$\fdg$6),
together with the very accurate {\it Chandra} localization, excludes 
the possibility of a misidentification. 

Searches in X-ray catalogues, and in the {\it ROSAT} all-sky survey 
(Voges et al. 1999) in particular, indicate that, despite its brightness 
in the 20--100 keV band, no high-energy data exist for this source below
20 keV. However, Revnivtsev et al. (2004b) and Sazonov \& Revnivtsev
(2004) report the existence of the {\it RXTE} source XSS J12389$-$1614. 
Albeit with large ($\sim$1$^{\circ}$) positional uncertainty, its 
localization is consistent with the hard X-ray emission seen with {\it 
INTEGRAL}. It has 3--8 keV and 8--20 keV fluxes of
(0.9$\pm$0.1)$\times$10$^{-11}$ erg cm$^{-2}$ s$^{-1}$ and
(1.0$\pm$0.2)$\times$10$^{-11}$ erg cm$^{-2}$ s$^{-1}$, respectively.

We moreover note that, in the {\it INTEGRAL} and {\it IRAS} error boxes of
IGR J12391$-$1610, a further radio emitter (labeled as NVSS 
J123911$-$161041) is found with a flux of 3.8$\pm$0.5 mJy at 1.4 GHz 
(Condon et al. 1998). Just outside the 3-$\sigma$ error circle of this 
radio source, the edge-on spiral galaxy (2MASX J12391039$-$1610432), 
located at $\sim$1$'$ from LEDA 170194, is present.
For this latter object, no information is available in the 
literature.

{\it IGR J18406$-$0539}: this source, with ISGRI coordinates RA = 
18$^{\rm h}$ 40$^{\rm m}$ 55$\fs$2, Dec = $-$05$^{\circ}$ 39$'$ 00$''$ 
(J2000), was detected at a (2.7$\pm$0.4)$\times$10$^{-11}$ erg cm$^{-2}$
s$^{-1}$ flux in the 18--60 keV band (Molkov et al. 2004). Within the 
{\it INTEGRAL} error box no peculiar catalogued X-ray or radio object is 
reported. However, at the southwestern edge of the error box, the optical 
emission-line star SS 406 is found (see Fig. 1, top right panel).
Stephenson \& Sanduleak (1977) classified SS 406 as a probable OBe star 
with weak H$_\alpha$ emission. The possible identification of this hard 
X-ray object as an early-type emission-line star suggests that SS 406 
could be the counterpart of IGR J18406$-$0539, in analogy with other HMXBs 
detected with INTEGRAL (e.g., Reig et al. 2005).
This star is present in the {\it Tycho} catalogue (H\o g et al. 2000)
with magnitudes $B_T$ = 12.857$\pm$0.309 mag and $V_T$ = 11.958$\pm$0.222 
mag. These, using appropriate conversion formulae (ESA 1997), correspond 
to a magnitude $V$ = 11.88$\pm$0.23 mag and to a color index 
$B-V$ = 0.76$\pm$0.32 mag in the Johnson system.

{\it 2E 1853.7+1534}: this object also has been detected in the
forthcoming 2$^{\rm nd}$ IBIS survey (Bassani et al. 2005; Bird et 
al. 2005) at coordinates RA = 18$^{\rm h}$ 56$^{\rm m}$ 01$\fs$9, 
Dec = +15$^{\circ}$ 37$'$ 16$''$ (J2000), with fluxes 
(2.1$\pm$0.2)$\times$10$^{-11}$ erg cm$^{-2}$ s$^{-1}$ (20--40 keV) 
and (1.9$\pm$0.4)$\times$10$^{-11}$ erg cm$^{-2}$ 
s$^{-1}$ (40--100 keV). This IBIS source is positionally coincident with 
an old {\it Einstein} detection (McDowell 1994), at a flux of 
$\sim$10$^{-12}$ erg cm$^{-2}$ s$^{-1}$ in the 0.16--3.5 keV band. 
It is also consistent with a {\it ROSAT} HRI/BMW object, 1BMW
J185600.5+153757 (Panzera et al. 2003), at a 0.1--2.0 keV flux
comparable to that of 2E 1853.7+1534. The coordinates (J2000) of this 
{\it ROSAT} object are RA = 18$^{\rm h}$ 56$^{\rm m}$ 00$\fs$48, 
Dec = +15$^{\circ}$ 37$'$ 56$\farcs$6, with a total error radius of 8$''$
once all systematic uncertainties are taken into account (Panzera et al. 
2003). A radio source, NVSS J185600+153755, with 1.4 GHz flux 
density of 3.4$\pm$0.4 mJy (Condon et 
al. 1998) and total positional uncertainty of 6$\farcs$3 is also
found in coincidence with the {\it ROSAT} position. The arcsec-sized {\it
ROSAT} and NVSS detections allowed us to identify 2E 1853.7+1534 with an
USNO-A2.0\footnote{available at \\ {\tt 
http://archive.eso.org/skycat/servers/usnoa}} object (Fig. 1, bottom 
left panel) having optical magnitudes $R \sim$ 
15.9 and $B \sim$ 18.4. These magnitudes indicate that this source is 
quite red ($B-R \sim$ 2.5 mag), possibly as a consequence of its location 
projectionally close to the Galactic Plane ($b$ = +6$\fdg$1). Its 
apparently extended shape, as seen on the DSS-II-Red 
Survey\footnote{available at \texttt{http://archive.eso.org/dss/dss/}}, 
points to an extragalactic origin for 2E 1853.7+1534.

{\it IGR J19473+4452}: it is one of the 8 objects included in the
{\it INTEGRAL}/{\it Chandra} minisurvey of Sazonov et al. (2005),
with ISGRI coordinates RA = 19$^{\rm h}$ 47$^{\rm m}$ 20$\fs$6, Dec = 
+44$^{\circ}$ 51$'$ 50$''$ (J2000). These authors report that this source 
has 0.5--8 keV and 17--60 keV fluxes of (3.0$\pm$1.0)$\times$10$^{-12}$ 
erg cm$^{-2}$ s$^{-1}$ and (2.5$\pm$0.4)$\times$10$^{-11}$ erg cm$^{-2}$ 
s$^{-1}$, respectively; moreover, a large neutral hydrogen column density, 
$N_{\rm H}$ = (11$\pm$1)$\times$10$^{22}$ cm$^{-2}$, appears to be present 
along the line of sight of this object according to their X-ray 
spectral data fitting. 
At the subarcsecond {\it Chandra} position, RA = 19$^{\rm h}$ 
47$^{\rm m}$ 19$\fs$37, Dec = +44$^{\circ}$ 49$'$ 42$\farcs$4 (J2000; 
error radius: $\sim$0$\farcs$6), a relatively bright optical and 
near-infrared object is 
detected (in Fig. 1, bottom right panel; see also Halpern 2005 and
Sazonov et al. 2005) with USNO-A2.0 magnitudes $R \sim$ 15.2 mag and 
$B \sim$ 15.7. This, despite the large $N_{\rm H}$ estimate above, 
indicates that the optical object is remarkably blue. Preliminary results
from Sazonov et al. (2005) show that this is an extragalactic object,
most likely an AGN, with redshift $z$ = 0.0539. No clearer indication on 
the exact nature of this source is however reported by these authors.

\begin{figure*}[t!]
\hspace{2cm}
\psfig{file=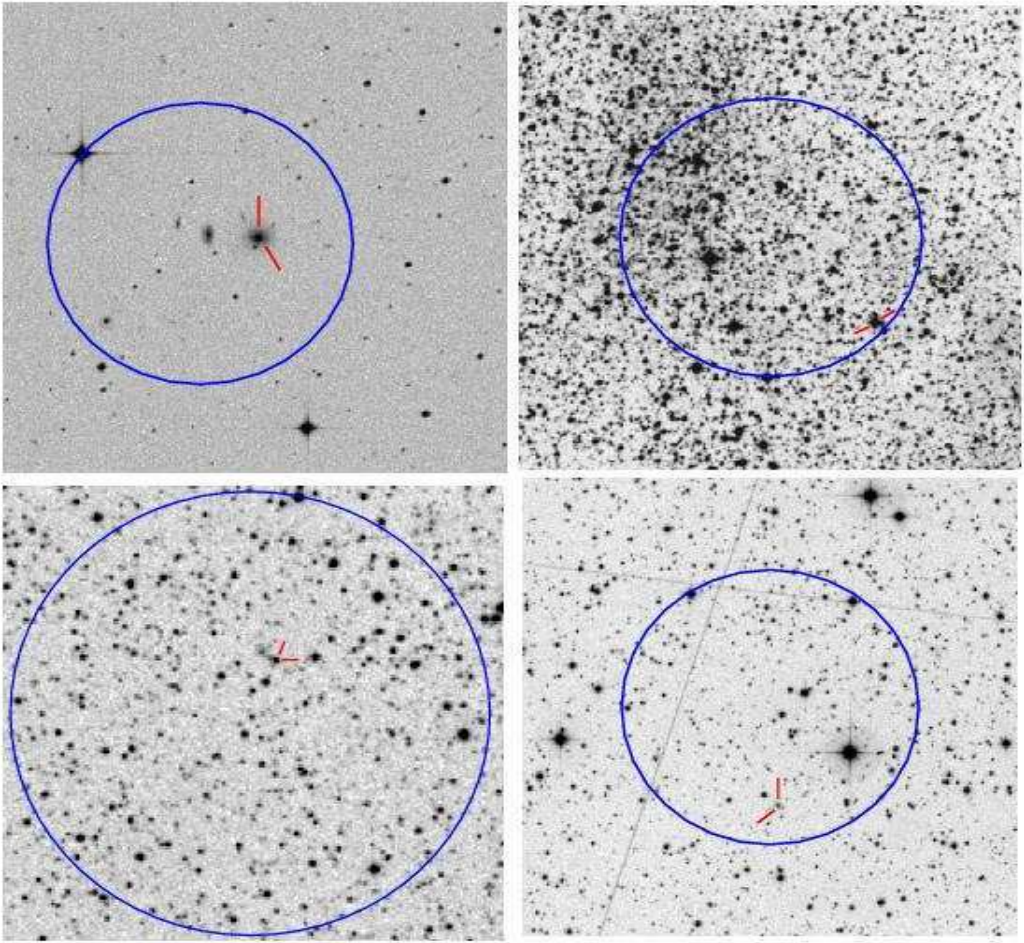,width=14cm,height=13cm}
\caption{DSS-II-Red optical images of the fields of IGR J12391$-$1610 
(top left panel), IGR J18406$-$0539 (top right panel), 2E 
1853.7+1534 (bottom left panel) and IGR J19473+4452 (bottom right 
panel). The putative optical counterparts are indicated with tick marks, 
while the circle mark the 3$'$ radius conservative ISGRI/{\it INTEGRAL} 
error boxes of the hard X-ray sources. Field sizes are 10$'$$\times$10$'$ 
for IGR J12391$-$1610, IGR J18406$-$0539 and IGR J19473+4452, and 
6$'$$\times$6$'$ for 2E 1853.7+1534. In all cases, North is up and East 
to the left. In the top left panel, the object 2MASX 
J12391039$-$1610432 is the edge-on galaxy located $\sim$1$'$ East of 
the galaxy LEDA 170194.}
\end{figure*}

\section{Optical observations in Loiano}

\begin{table*}[t!]
\caption[]{Log of the spectroscopic observations presented in this paper.}
\begin{center}
\begin{tabular}{llcccc}
\noalign{\smallskip}
\hline
\noalign{\smallskip}
\multicolumn{1}{c}{Object} & \multicolumn{1}{c}{Date} & Mid-exposure & 
Grism & Slit & Exposure \\
 & & time (UT) & number & (arcsec) & time (s) \\
\noalign{\smallskip}
\hline
\noalign{\smallskip}

IGR J12391$-$1610 (=LEDA 170194) & 01 Apr 2005 & 22:45:46 & \#4 & 2.0 & 2400 \\

IGR J18406$-$0539 (=SS 406) & 06 Jun 2005 & 23:51:25 & \#4 & 2.0 & 2$\times$600 \\

2E 1853.7+1534    & 06 Jun 2005 & 22:43:47 & \#4 & 2.0 & 1800 \\

IGR J19473+4452  & 01 Sep 2005 & 21:45:59 & \#4 & 2.0 & 1800 \\

\noalign{\smallskip}
\hline
\noalign{\smallskip}
\end{tabular}
\end{center}
\end{table*}

The Bologna Astronomical Observatory 1.52-metre ``G.D. Cassini'' telescope
plus BFOSC was used to spectroscopically observe the galaxy LEDA 170194,
the OBe star SS 406, and the putative optical counterparts to the 
{\it INTEGRAL} sources 2E 1853.7+1534 and IGR J19473+4452 
(see Fig. 1). The BFOSC instrument is
equipped with a 1300$\times$1340 pixel EEV CCD. In all observations,
Grism \#4 and a slit width of $2''$ were used, providing a 3500--8500
\AA~nominal spectral coverage. The use of this setup secured a final
dispersion of 4.0~\AA/pix for all spectra. The spectrum of LEDA 170194 was 
acquired in such a way that the slit also included the closeby galaxy 
2MASX J12391039$-$1610432. All observations
were performed with the slit in the E-W direction; this, in 
particular for LEDA 170194 which was observed at large airmass ($\sim$2),
may induce nonperfect flux calibration at the blue edge of the spectrum 
(that is, bluewards of 4000 \AA) due to the fact that the slit was 
not oriented along the parallactic angle. The complete log of the 
observations is reported in Table 1.

The spectra, after cosmic-ray rejection, were reduced, background
subtracted and optimally extracted (Horne 1986) using IRAF\footnote{IRAF
is the Image Analysis and Reduction Facility made available to the
astronomical community by the National Optical Astronomy Observatories,
which are operated by AURA, Inc., under contract with the U.S. National
Science Foundation. It is available at {\tt http://iraf.noao.edu/}}.
Wavelength calibration was performed using He-Ar lamps acquired soon 
after each spectroscopic exposure; all spectra were then flux-calibrated 
by applying a library response function built using the spectrophotometric 
standard BD+25$^\circ$3941 (Stone 1977). When applicable, different 
spectra of the same object were stacked together to increase the S/N 
ratio. Wavelength calibration uncertainty was $\sim$0.5~\AA~for all cases; 
this was checked by using the positions of background night sky lines.

\begin{figure*}[t!]
\centering{\mbox{\psfig{file=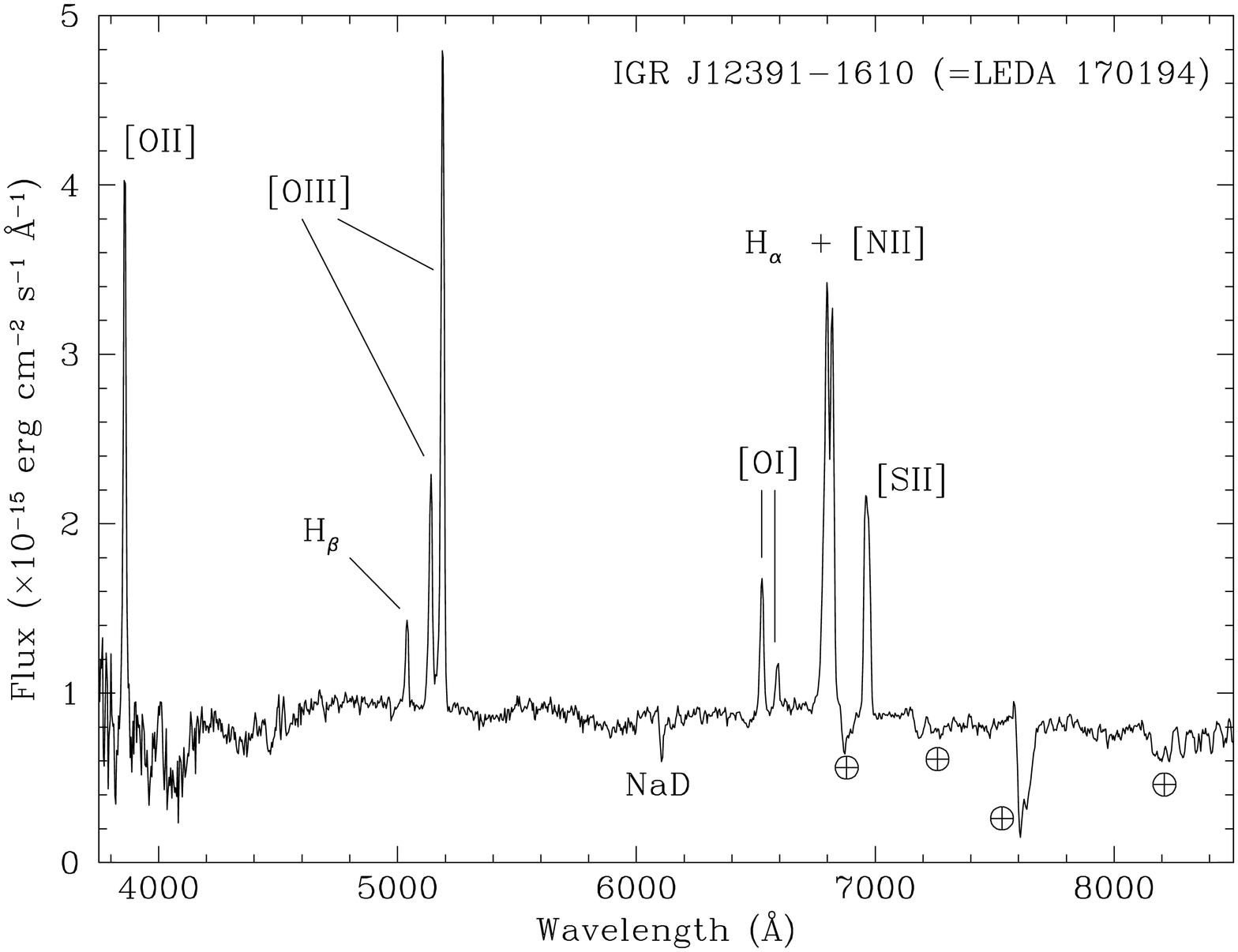,width=8.9cm,angle=270}}}
\centering{\mbox{\psfig{file=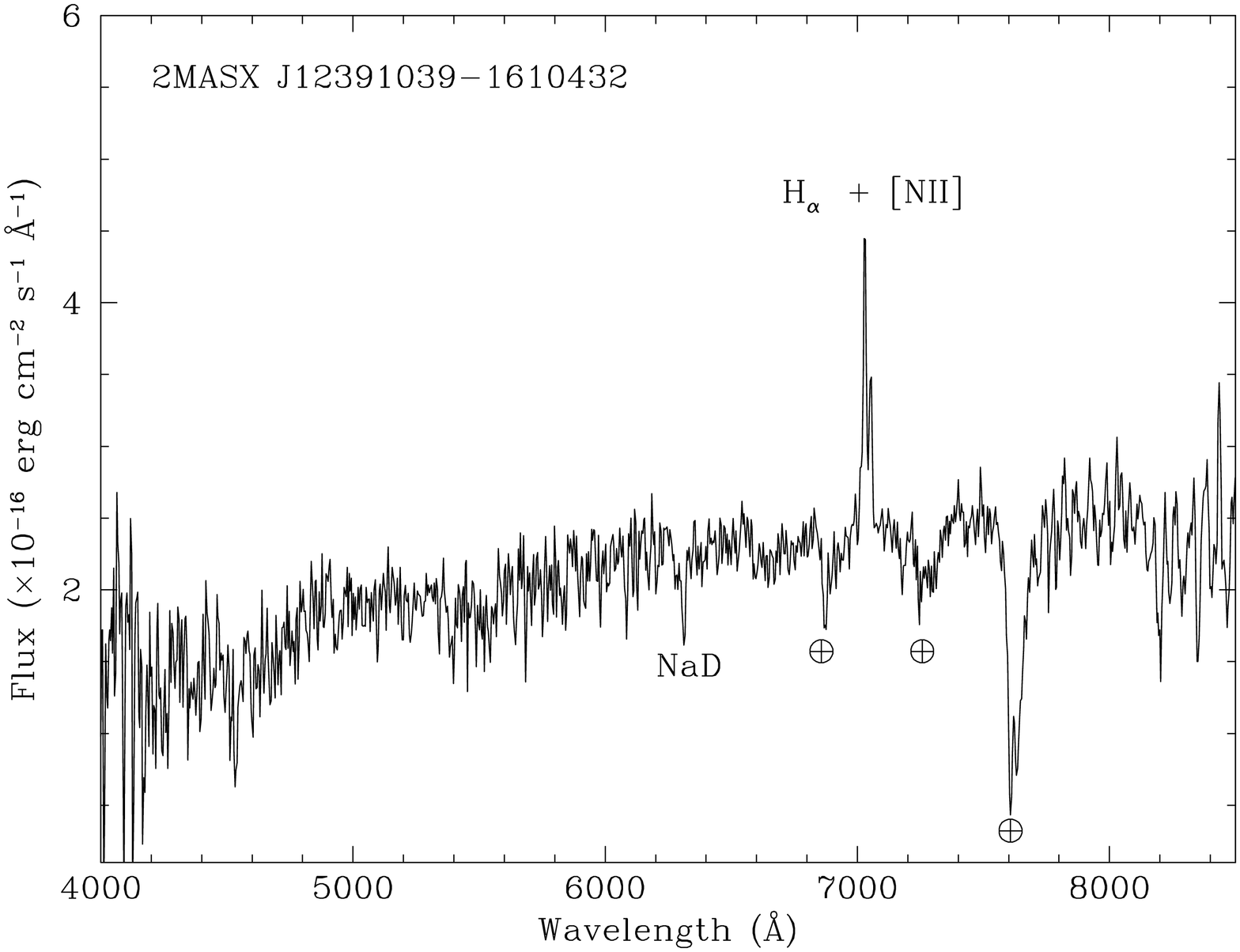,width=8.9cm,angle=270}}}
\centering{\mbox{\psfig{file=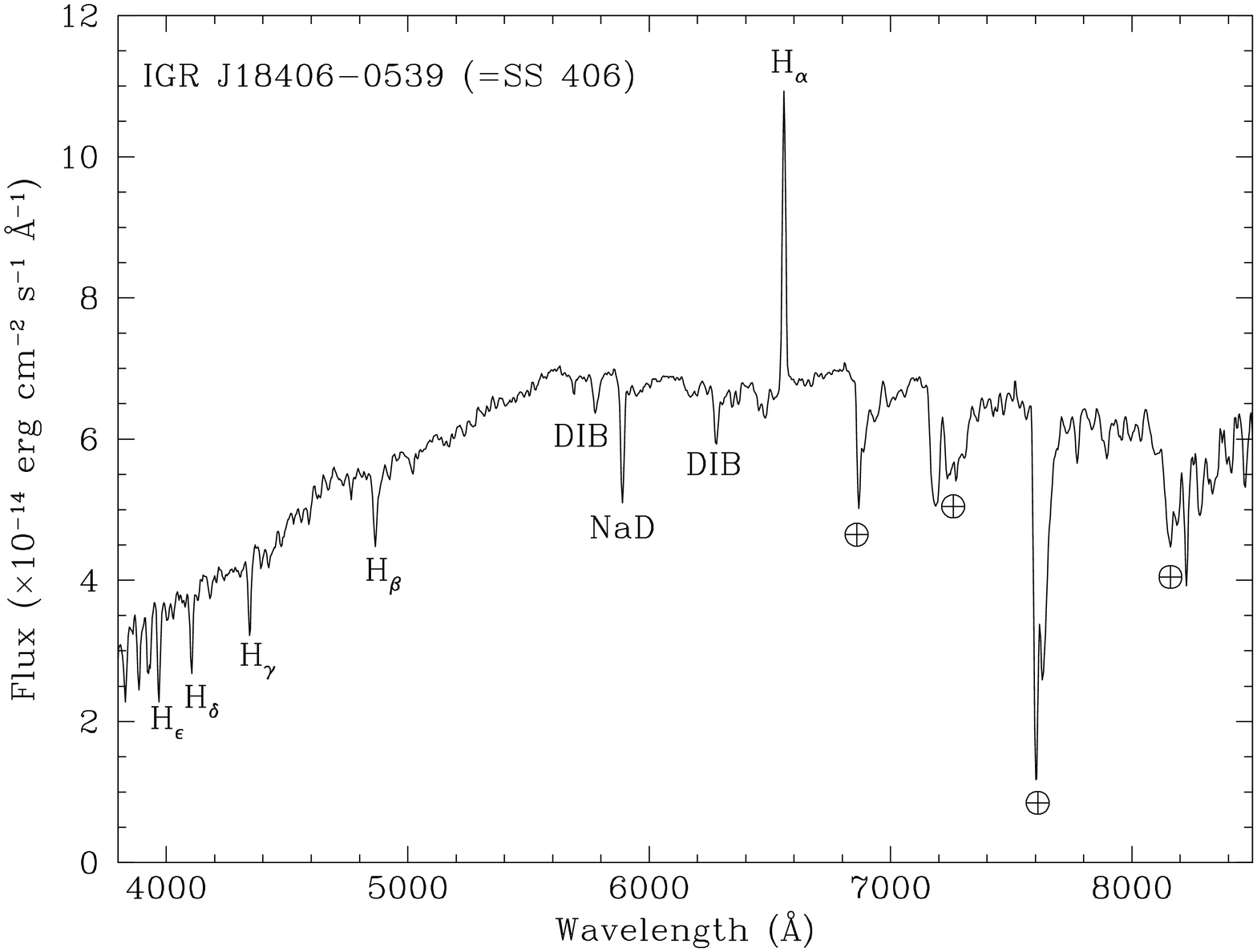,width=8.9cm,angle=270}}}
\centering{\mbox{\psfig{file=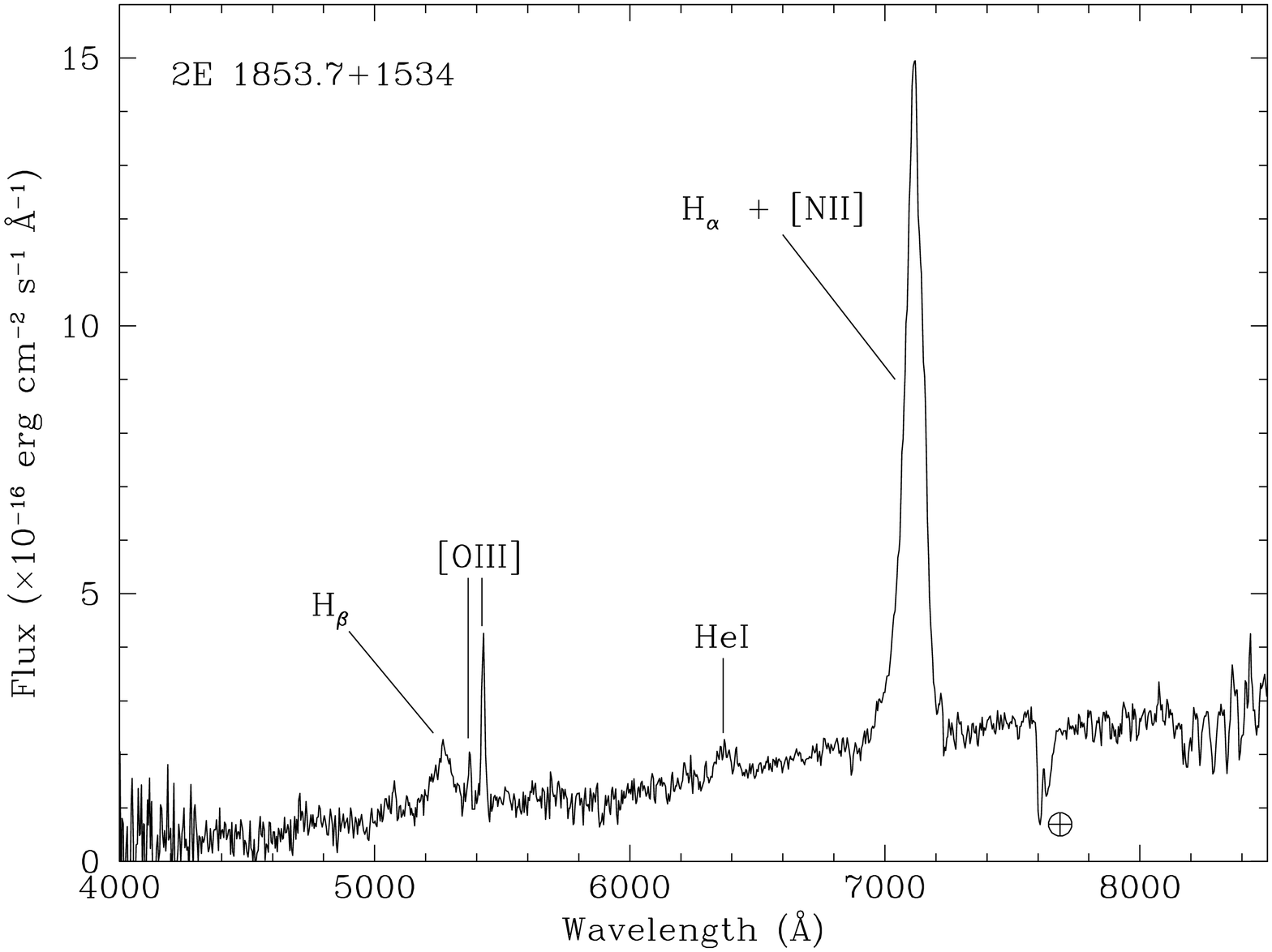,width=8.9cm,angle=270}}}
\begin{center}
\parbox{9cm}{
\hspace{-.2cm}
\psfig{file=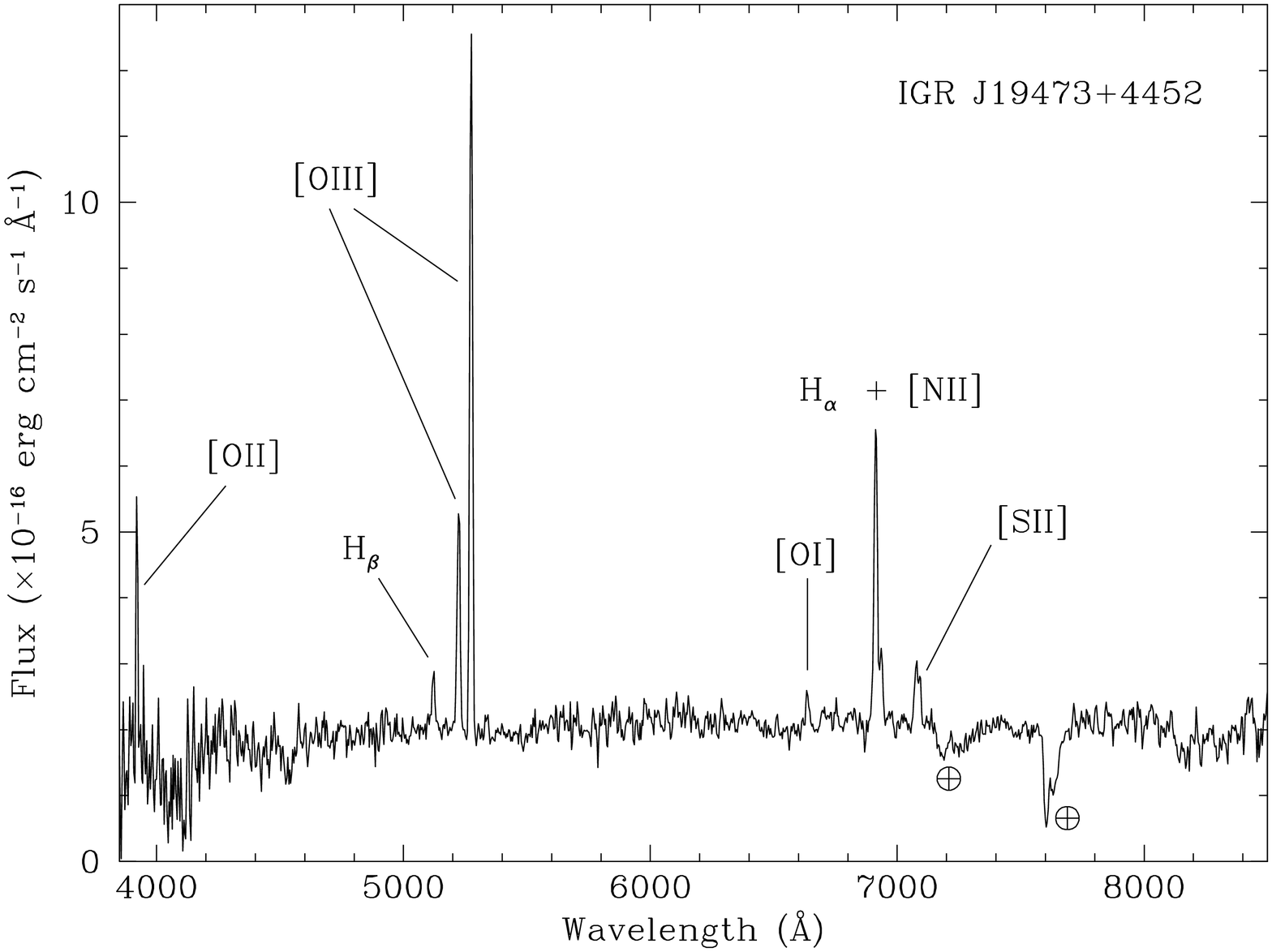,width=8.9cm,angle=270}
}
\parbox{8.5cm}{
\vspace{-1cm}
\caption{Spectra (not corrected for the intervening Galactic absorption) 
of the optical counterparts to IGR J12391$-$1610 (=LEDA 
170194; upper left panel), IGR J18406$-$0539 (= SS 406; central left 
panel), 2E 1853.7+1534 (central right panel) and IGR J19473+4452
(lower left panel) acquired with the Cassini 
telescope at Loiano. The spectrum of the galaxy 2MASX J12391039$-$1610432
in the field of IGR J12391$-$1610 is also reported 
(upper right panel). For each spectrum the main spectral features are labeled. 
The symbol $\oplus$ indicates atmospheric telluric features.}
}
\end{center}
\end{figure*}

\section{Results and discussion}

Table 2 reports the (observer's frame) emission-line wavelengths, fluxes
and equivalent widths (EWs) of the five observed objects reported in 
Fig. 2. The line fluxes from extragalactic objects were dereddened for 
Galactic
absorption along the respective lines of sight following the prescription
of Schlegel et al. (1998; see below). These same spectra were also not 
corrected for starlight contamination (see, e.g., Ho et al. 1993, 1997)  
given the limited S/N and resolution of the spectrum; however, we do not 
expect that this will affect any of our conclusions.
Moreover, in the following we assume a cosmology with $H_{\rm 0}$ = 65 km 
s$^{-1}$ Mpc$^{-1}$, $\Omega_{\Lambda}$ = 0.7 and $\Omega_{\rm m}$ = 0.3
(e.g., Koopmans \& Fassnacht 1999).

\subsection{IGR J12391$-$1610 (=LEDA 170194)}

The spectra of the galaxy LEDA 170194 (Fig. 2, upper left) shows a number
of narrow emission features that can be readily identified with redshifted
optical nebular lines. These include [O~{\sc ii}] $\lambda$3727, H$_\beta$, 
[O~{\sc iii}] $\lambda\lambda$4958,5007, H$_\alpha$, [N~{\sc ii}]
$\lambda\lambda$6548,6583, and [S~{\sc ii}] $\lambda\lambda$6716,6731.
All identified emission lines yield a redshift of $z$ = 0.036$\pm$0.001, 
in perfect agreement with da Costa et al. (1998). The NaD doublet in 
absorption is also detected at the same redshift.

The presence of just narrow emission lines in the optical spectrum of 
LEDA 170194 indicates that they are due to the activity of an obscured
AGN; this is also suggested by the NVSS radio detection and by the {\it
ROSAT} nondetection in soft X-rays. We further confirm this by using
the diagnostic line ratios [N~{\sc ii}]/H$_\alpha$ (= 0.95$\pm$0.04), 
[S~{\sc ii}]/H$_\alpha$ (= 0.86$\pm$0.04), and [O~{\sc iii}]/H$_\beta$
(= 8.7$\pm$0.7), together with the detection of substantial 
[O~{\sc i}] $\lambda$6300 emission: the values of these parameters place 
this source in the regime of Seyfert 2 AGNs (Ho et al. 1993, 1997).

Using the cosmology described above and the more accurate redshift
of da Costa et al. (1998) we find that the luminosity distance
to the galaxy LEDA 170194 is $d_L$ = 174 Mpc, and that its X-ray
luminosities are 7.4$\times$10$^{42}$~erg s$^{-1}$ and
2.6$\times$10$^{44}$~erg s$^{-1}$ in the 0.5--8 keV and 
20--100 keV bands, respectively.
These values place the source among the most luminous Type 2 Seyfert
galaxies detected so far (e.g., Risaliti 2002; Sazonov \& Revnivtsev 
2004). The measured values for the X-ray luminosities of LEDA 170194 
are thus comparable with that of ``classical'' AGNs.

The strength of the optical emission lines of LEDA 170194, after 
accounting for Galactic and intrinsic absorptions, can be used to 
estimate the star formation rate (SFR) and metallicity. First, a
correction for Galactic reddening has been applied (we assumed a color 
excess $E(B-V)$ = 0.046 mag following Schlegel et al. 1998). Next, 
considering an intrinsic Balmer decrement of H$_\alpha$/H$_\beta$ = 2.86 
(Osterbrock 1989) and the extinction law of Cardelli et al. (1989), the 
observed flux ratio H$_\alpha$/H$_\beta$ = 6.4 implies an internal color 
excess $E(B-V)$ = 0.80 mag (in the galaxy rest frame). Following Kennicutt 
(1998), we determine a SFR of 10$\pm$1 $M_\odot$ yr$^{-1}$ from the
reddening-corrected H$_\alpha$ luminosity of (1.20$\pm$0.08)$\times$10$^{42}$ 
erg s$^{-1}$. The method (again in Kennicutt 1998) which instead uses 
the extinction-corrected [O {\sc ii}] luminosity, 
(8.0$\pm$0.7)$\times$10$^{42}$ erg s$^{-1}$, yields
a much larger SFR value, 110$\pm$30 $M_\odot$ yr$^{-1}$. This
high value, so different from that obtained using the H$_\alpha$ emission, 
may be produced by the abovementioned uncertainty in the flux 
calibration of the spectrum at its blue edge (see Sect. 3), so we should 
treat this latter SFR estimate very cautiously.
Moreover, as stressed by Kennicutt (1998), the SFR determination from the 
[O {\sc ii}] line flux suffers from larger uncertainties with respect to 
that obtained from the H$_\alpha$ emission line.

The total reddening estimate along the line of sight inferred from 
the optical spectrum corresponds, using the empirical formula of Predehl 
\& Schmitt (1995), to a $N_{\rm H} \approx$ 5$\times$10$^{21}$ cm$^{-2}$, 
which is $\sim$4 times less than that measured by Sazonov et al. 
(2005) from {\it Chandra} X-ray data.

Next, assuming for IGR J12391$-$1610 the best-fit X-ray spectrum as in
Sazonov et al. (2005), we can determine the 2--10 keV flux of the source. 
This results in 2.2$\times$10$^{-12}$ erg cm$^{-2}$ s$^{-1}$. The 
comparison between the reddening-corrected [O~{\sc iii}] $\lambda$5007 
emission flux and the 2--10 keV X-ray flux estimated above implies an 
X-ray/[O~{\sc iii}]$_{\rm 5007}$ ratio of $\sim$2.5, which 
indicates that this source is in the Compton-thick regime 
(see Bassani et al. 1999).

Although, as we said above, the [O {\sc ii}] emission line flux estimate
is affected by large uncertainties, the detection of [O {\sc ii}], [O {\sc
iii}], and H$_\beta$ also allows us to infer the gaseous oxygen abundance
in this galaxy. Following Kobulnicky et al. (1999), the $R_{\rm 23}$
parameter, defined as the ratio between [O {\sc ii}] + [O {\sc iii}] and
H$_\beta$ line fluxes, gives 12 + log (O/H) $\approx$ 8.5. Considering the
intrinsic luminosity of the source (it has rest-frame absolute $B$-band
magnitude M$_B$ = $-$21.27 mag; Prugniel 2005) and its [O {\sc
iii}]/[N~{\sc ii}] ratio ($\sim$3), all of this information points to a
basically solar oxygen abundance. A similar result is obtained if we use
the [N~{\sc ii}]/H$_\alpha$ flux ratio method (Kewley \& Dopita 2002).

As mentioned in Sect. 3, we also acquired a spectrum of 2MASX 
J12391039$-$1610432, the edge-on galaxy located 1$'$ east 
of LEDA 170194. The spectrum, albeit noisy (see Fig. 2, upper right), shows
the presence of prominent and narrow H$_\alpha$ and [N~{\sc ii}]
$\lambda\lambda$6548,6583 emission lines at redshift $z$ =
0.071$\pm$0.001. In this case also, the NaD doublet in absorption is 
detected at this same redshift. This implies a luminosity distance 
$d_L$ = 345 Mpc for this galaxy, thus twice as far from Earth as 
LEDA 170194.

The detection of [N {\sc ii}] and H$_\alpha$ in the spectrum of 2MASX 
J12391039$-$1610432 also allows us to infer the gaseous oxygen abundance 
in this galaxy. The use of the [N {\sc ii}]/H$_\alpha$ ratio method (among 
those listed in Kewley \& Dopita 2002) for the determination of the 
metallicity of this galaxy is indicated because it is the least sensitive 
to, and therefore not substantially influenced by, the lack of our 
knowledge of the absorption intrinsic to this galaxy. 
Indeed, the two emission lines are so close to each other that the 
differential intrinsic reddening is not significant. We thus find 
that the [N {\sc ii}]/H$_\alpha$ ratio observed here implies 12 + log 
(O/H) $\approx$ 9. Therefore, in this case also we find an oxygen 
abundance which is consistent with the solar value.

The intensity of the H$_\alpha$ emission line of 2MASX J12391039$-$1610432,
once corrected for the Galactic absorption, can also be used to estimate
the SFR in this galaxy. Using Eq. (2) of Kennicutt (1998) we determine a 
SFR of 1.40$\pm$0.15 $M_\odot$ yr$^{-1}$. This should conservatively be 
considered as a lower limit to the SFR because the effect of absorption 
intrinsic to 2MASX J12391039$-$1610432 was not accounted for.

The poor S/N of the spectrum of this source does not allow us to deduce 
much more about the nature of this narrow H$_\alpha$ emission-line galaxy. 
We of course can exclude that it is a Seyfert 1 type AGN due to the 
absence of broad emission lines. But, given its optical (and maybe 
radio) activity, we cannot {\it a priori} exclude that this object is 
co-responsible, together with LEDA
170194, for the X-ray emission detected by {\it INTEGRAL} as IGR
J12391$-$1610. However, a quick look at a 3.3 ks Chandra observation (Seq. 
Num.: 701178, Obs. ID: 6276, PI: R.A. Sunyaev) acquired on July 25, 2005,
does not show detectable X-ray emission either at the 2MASX 
J12391039$-$1610432 position or within the NVSS J123911$-$161041 radio 
error circle, whereas X-rays are clearly detected from LEDA 170194
(see also Halpern 2005 and Sazonov et al. 2005).
This implies that either the soft ($<$10 keV) X-ray emission, if any, 
from the former sources is heavily absorbed or, more likely, that they
are not X-ray emitting and that LEDA 170194 is solely responsible for 
the hard X-rays detected by {\it INTEGRAL}. In this latter case, the 
galaxy 2MASX J12391039$-$1610432 can be identified as a 
starburst/H{\sc ii} galaxy. 

Regarding the association between this galaxy and the nearby 
NVSS radio source, we note that, given the relatively low 
S/N ratio of the radio detection, the NVSS position and the 
corresponding uncertainties may not be very accurate; so, they might 
actually be positionally consistent with each other. Thus, only 
detailed radio observations can give an answer to this open 
issue.

\begin{table}[t!]
\caption[]{Observer's frame wavelengths, EWs (both in \AA ngstroms) and
fluxes (in units of 10$^{-15}$ erg s$^{-1}$ cm$^{-2}$) of the emission
lines detected in the spectra of the five objects reported in Fig. 2. 
For the extragalactic objects the values are corrected for Galactic reddening
assuming (from Schlegel et al. 1998) $E(B-V)$ = 0.046 mag along the LEDA
170194 and 2MASX J12391039$-$1610432 line of sight, $E(B-V)$ = 0.94 
mag along the 2E 1853.7+1534 line of sight and $E(B-V)$ = 0.20 
mag along the IGR J19473+4452 line of sight. The error on the line 
positions is conservatively assumed to be $\pm$4 \AA, i.e., comparable 
with the spectral dispersion (see text).}
\begin{center}
\begin{tabular}{lcrr}
\noalign{\smallskip}
\hline
\hline
\noalign{\smallskip}
\multicolumn{1}{c}{Line} & $\lambda_{\rm obs}$ (\AA) & 
\multicolumn{1}{c}{EW$_{\rm obs}$ (\AA)} & \multicolumn{1}{c}{Flux} \\
\noalign{\smallskip}
\hline
\noalign{\smallskip}

\multicolumn{4}{c}{IGR J12391$-$1610 (=LEDA 170194)} \\

$[$O {\sc ii}$]$ $\lambda$3727  & 3858 &   85$\pm$10  &   64$\pm$4   \\
H$_\beta$                       & 5039 &  8.6$\pm$0.6 &  7.9$\pm$0.6 \\
$[$O {\sc iii}$]$ $\lambda$4958 & 5139 & 26.5$\pm$1.3 & 24.3$\pm$1.2 \\
$[$O {\sc iii}$]$ $\lambda$5007 & 5189 &   75$\pm$2   &   69$\pm$2   \\
$[$O {\sc i}$]$ $\lambda$6300   & 6526 & 16.8$\pm$0.8 & 17.3$\pm$0.9 \\
$[$O {\sc i}$]$ $\lambda$6363   & 6591 &  6.2$\pm$0.4 &  6.2$\pm$0.4 \\
$[$N {\sc ii}$]$ $\lambda$6548  & 6783 & 20.4$\pm$1.4 & 20.7$\pm$1.4 \\
H$_\alpha$                      & 6798 & 49.1$\pm$1.5 & 50.2$\pm$1.5 \\
$[$N {\sc ii}$]$ $\lambda$6583  & 6822 & 46.5$\pm$1.4 & 47.7$\pm$1.4 \\
$[$S {\sc ii}$]$ $\lambda$6716  & 6959 & 22.6$\pm$1.1 & 23.6$\pm$1.2 \\
$[$S {\sc ii}$]$ $\lambda$6731  & 6973 & 19.0$\pm$1.0 & 19.8$\pm$1.0 \\

\noalign{\smallskip}
\hline
\noalign{\smallskip}

\multicolumn{4}{c}{2MASX J12391039$-$1610432} \\

$[$N {\sc ii}$]$ $\lambda$6548  & 7013 & 0.53$\pm$0.13 & 2.0$\pm$0.5 \\
H$_\alpha$                      & 7030 &  3.3$\pm$0.3  & 12.5$\pm$1.3 \\
$[$N {\sc ii}$]$ $\lambda$6583  & 7052 &  1.9$\pm$0.3  & 7.1$\pm$1.1  \\

\noalign{\smallskip}
\hline
\noalign{\smallskip}

\multicolumn{4}{c}{IGR J18406$-$0539 (=SS 406)} \\

H$_\alpha$                      & 6559 & 10.6$\pm$0.2 & 680$\pm$30 \\

\noalign{\smallskip}
\hline
\noalign{\smallskip}

\multicolumn{4}{c}{2E 1853.7+1534} \\

H$_\beta$                         & 5268 &  89$\pm$9  & 161$\pm$17 \\
$[$O {\sc iii}$]$ $\lambda$4958   & 5372 &  11$\pm$3  &  20$\pm$5  \\
$[$O {\sc iii}$]$ $\lambda$5007   & 5425 &  44$\pm$7  & 75$\pm$11  \\
He {\sc i} $\lambda$5875	  & 6368 &  26$\pm$5  & 41$\pm$8   \\
H$_\alpha$ + $[$N {\sc ii}$]$$^*$ & 7113 & 470$\pm$20 & 790$\pm$40 \\

\noalign{\smallskip}
\hline
\noalign{\smallskip}

\multicolumn{4}{c}{IGR J19473+4452} \\

$[$O {\sc ii}$]$ $\lambda$3727  & 3920 & 23$\pm$6     &  8$\pm$2       \\
H$_\beta$                       & 5121 &  6.6$\pm$1.3 &  2.4$\pm$0.5   \\
$[$O {\sc iii}$]$ $\lambda$4958 & 5224 & 25$\pm$3     &  9.0$\pm$0.9   \\
$[$O {\sc iii}$]$ $\lambda$5007 & 5274 & 76$\pm$4     & 26.8$\pm$1.4   \\
$[$O {\sc i}$]$ $\lambda$6300   & 6636 &  3.5$\pm$0.9 &  1.1$\pm$0.3   \\
H$_\alpha$                      & 6913 & 35$\pm$4     & 11.2$\pm$0.8   \\
$[$N {\sc ii}$]$ $\lambda$6583  & 6936 &  7.3$\pm$0.7 &  2.3$\pm$0.2   \\
$[$S {\sc ii}$]$ $\lambda$6716  & 7077 &  7.6$\pm$0.8 &  2.4$\pm$0.2   \\
$[$S {\sc ii}$]$ $\lambda$6731  & 7093 &  3.2$\pm$0.5 &  1.01$\pm$0.15 \\

\noalign{\smallskip}
\hline
\noalign{\smallskip}
\multicolumn{4}{l}{$^*$: these lines are heavily blended. The wavelength} \\
\multicolumn{4}{l}{of the emission peak is reported.} \\
\end{tabular}
\end{center}
\end{table}

\subsection{IGR J18406$-$0539 (=SS 406)}

The optical spectrum of SS 406 is reported in the central left panel 
of Fig. 2. The absence of He {\sc ii} lines points to a B-star 
classification for this object. Moreover, the shape of the Balmer 
absorption lines, along with the detection of fainter absorption 
features produced by He {\sc i} $\lambda\lambda$4026,4471 and by 
light metals (such as Si {\sc ii} $\lambda$4128, C {\sc ii} $\lambda$4267 
and Mg {\sc ii} $\lambda$4481), points to a main-sequence, mid-type B 
star (most likely B5) identification. Finally, the presence of a 
strong H$_\alpha$ line in emission (possibly showing a P-Cyg profile) 
allows us conclusively to classify SS 406 as a Be star.

Assuming no absorption along the line of sight, a spectral type B5V 
for SS 406 (which implies an absolute magnitude M$_V$ = $-$1.2 mag; 
Jaschek \& Jaschek 1987) and using the observed $V$-band magnitude  
$V$ = 11.88$\pm$0.23 mag (See Sect. 1.2), one obtains that the distance 
to the source is $d \sim$ 4 kpc. This should be considered as an upper 
limit, as no correction for Galactic absorption was taken into account. 
However, we expect that significant reddening is present towards
SS 406, given its Galactic latitude ($b$ = $-$0$\fdg$2), the shape of 
its observed spectral continuum, the total EW (4.5$\pm$0.2 \AA) of the 
Na Doublet at 5890 \AA, and the presence of other absorption features 
which are due to interstellar matter (see Fig. 2, central left panel). 

A more accurate estimate for the distance can be obtained by considering
the intrinsic and observed $B-V$ color indices of the star, i.e. 
$(B-V)_0$ = $-$0.15 mag (Wegner 1994) and $B-V$ = 0.76$\pm$0.32 mag, 
respectively. Their difference implies a color excess $E(B-V)$ = 
0.91$\pm$0.32 for SS 406 in the hypothesis that no further 
emission from the accreting object contributes to the total optical
light. By correcting for this color excess using the reddening law
of Cardelli et al. (1989), we get an apparent unabsorbed
$V$-band magnitude $V_0$ = 9.0$\pm$1.0 mag, which in turn 
gives a distance $d$ = 1.1$^{+0.6}_{-0.4}$ kpc assuming the 
absolute $V$ magnitude reported above. This distance is marginally 
compatible with SS 406 being located in the Sagittarius Arm (which lies at 
$\sim$2 kpc; see, e.g., Molkov et al. 2004), and implies an 18--60 keV 
luminosity of $\sim$4$\times$10$^{33}$ erg s$^{-1}$. 
This, together with the EW of the H$_\alpha$ emission line, is typical of 
low-luminosity, persistently-emitting Galactic HMXBs (see e.g. White et 
al. 1995).

We note that IGR J18406$-$0539 is located $\sim$4$'$ away (and 
not 7$'$ as reported in Rodriguez et al. 2004)
with respect to the other {\it INTEGRAL}/{\it ASCA} transient source IGR
J18410$-$0535/AX J1841.0$-$0536, thus marginally consistent with it
(Halpern et al. 2004; Bamba et al. 2001; Rodriguez et al. 2004). However,
the fact that the refined {\it Chandra} position and the optical
counterpart to IGR J18410$-$0535 (Halpern \& Gotthelf 2004) lie 
3$\farcm$5 from the IGR J18406$-$0539 position, thus formally outside its 
error box (and moreover are 6$\farcm$1 away from SS 406), 
suggests that these two {\it INTEGRAL} sources are not the
same. Besides, assuming an average number of $\sim$0.05 Be stars per
arcmin$^2$ along the Galactic Plane (see Reig et al. 2005), we find that
the chance probability of observing two Be stars within a radius of
$\sim$3$'$ is around 7\%. Thus, although we cannot exclude that IGR
J18406$-$0539 and IGR J18410$-$0535 are the same source (in this case, the
detection of the former actually corresponds to the quiescent state of the
latter) we put forward that, if the two {\it INTEGRAL} sources are
independent, we regard the association between IGR J18406$-$0539 and SS
406 as likely.

\subsection{2E 1853.7+1534}

In the spectrum of 2E 1853.7+1534 (in Fig. 2, centre right) the
most striking spectral feature is a prominent and broad redshifted
H$_\alpha$+[N {\sc ii}] emission blend. A broad H$_\beta$ emission line,
as well as [O~{\sc iii}] $\lambda$5007 narrow forbidden lines and possibly
a broad He~{\sc i} $\lambda$5875 emission are also detected. All of these
features have a redshift $z$ = 0.084$\pm$0.001. The presence of these 
emissions imply that this source is a Type 1 Seyfert galaxy according to, 
e.g., the classification of Osterbrock (1989).

Assuming the cosmology described above, this redshift means a luminosity 
distance of 412 Mpc for 2E 1853.7+1534 and X-ray luminosities of
2.0$\times$10$^{43}$ erg s$^{-1}$ and 8.1$\times$10$^{44}$ erg 
s$^{-1}$ in the 0.1--2 keV and 20--100 keV bands, respectively.
Analogously, this distance implies an absolute optical $B$-band magnitude
M$_B \sim$ $-$23.5 mag. This is, strictly speaking, an actual lower limit 
to the $B$-band luminosity of 2E 1853.7+1534, as no absorption internal to
the AGN host galaxy was considered. However, substantial intrinsic 
reddening is not expected in Seyfert 1 galaxies, so we can confidently 
consider this value for M$_B$ as close to the real one. All of 
these luminosity estimates place 2E 1853.7+1534 at the bright end of the 
Seyfert 1 galaxies distribution (Perola et al. 2002).

Next, following Kaspi et al. (2000) and Wu et al. (2004), we can compute
an estimate of the mass of the central black hole in this active galaxy.
This can be achieved using (i) the flux of the H$_\beta$ emission
(in Table 2),
corrected considering a foreground Galactic color excess $E(B-V)$ = 0.94
(Schlegel et al. 1998) and (ii) a broad-line region gas velocity $v_{\rm
BLR} \sim$ ($\sqrt{3}$/2)$\cdot$$v_{\rm FWHM}$ $\sim$ 4200 km 
s$^{-1}$ (where $v_{\rm FWHM}$ $\sim$ 4800 km s$^{-1}$ is the 
velocity measured from the FWHM of the H$_\beta$ emission line).
From Eq. (2) of Wu et al. (2004) we find that the BLR size is $R_{\rm BLR} 
\sim$ 54 light-days. Furthermore, using Eq. (5) of Kaspi et al. (2000), 
the AGN black hole mass in 2E 1853.7+1534 is $M_{\rm BH} \sim$ 
1.4$\times$10$^{8}$ $M_\odot$.
Again, this is a lower limit (but likely close to the real value 
for the reasons explained above) as no absorption intrinsic to the 
AGN was accounted for.

\subsection{IGR J19473+4452}

Analogously to the case of the LEDA 170194 (Sect. 4.1), the spectrum of 
the putative counterpart to IGR J19473+4452 (Fig. 2, lower left) shows 
several narrow emission lines, which we identified as [O~{\sc ii}] 
$\lambda$3727, H$_\beta$, [O~{\sc iii}] $\lambda\lambda$4958,5007, 
H$_\alpha$, [N~{\sc ii}] $\lambda$6583, and [S~{\sc ii}] 
$\lambda\lambda$6716,6731. All of these emission features lie at a 
redshift $z$ = 0.053$\pm$0.001, consistent with Sazonov et al. (2005).

In this case also, the exclusive presence of narrow emission lines in the 
spectrum of the optical counterpart to IGR J19473+4452 points to the fact 
that they originate within a Narrow-Line Region of an AGN. A confirmation 
of this comes by examining the diagnostic line ratios of Ho et al. (1993, 
1997). These, [N~{\sc ii}]/H$_\alpha$ = 0.21$\pm$0.03, [S~{\sc 
ii}]/H$_\alpha$ = 0.30$\pm$0.06 and [O~{\sc iii}]/H$_\beta$ = 
11.2$\pm$2.3, place IGR J19473+4452 among Seyfert 2 AGNs.

To compute the internal reddening of this galaxy, we again use the procedure 
described in Sect. 3.1. We find that the
H$_\alpha$/H$_\beta$ flux ratio, once corrected for the Galactic
absorption $E(B-V)$ = 0.20 mag (according to Schlegel et al. 1998),
is 4.59; this indicates a rest-frame internal color excess $E(B-V)$ = 0.48 
mag for IGR J19473+4452.
We note that the total reddening estimate along the 
line of sight corresponds, using the empirical formula of Predehl \& 
Schmitt (1995), to a neutral hydrogen column density $N_{\rm H} \approx$ 
4$\times$10$^{21}$ cm$^{-2}$, that is $\sim$30 times less than the 
$N_{\rm H}$ measure obtained by Sazonov et al. (2005) from {\it Chandra} 
X-ray data.

The measured redshift implies a luminosity distance to this source of 254 
Mpc, and thus X-ray luminosities of 2.3$\times$10$^{43}$~erg s$^{-1}$ 
and 1.9$\times$10$^{44}$~erg s$^{-1}$ in the 0.5--8 keV and 17--60 
keV bands, respectively. Using the $B$-band optical magnitude of this 
object, the above distance points to an absolute $B$ magnitude M$_B \sim$ 
$-$23.4 mag. These values place this source in the bright side of 
the Type 2 Seyfert galaxies luminosity distribution (Risaliti 2002; Sazonov 
\& Revnivtsev 2004).

In the same way as performed for IGR J12391$-$1610, we can determine the 
Compton regime for IGR J19473+4452. Using the X-ray spectral information 
of Sazonov et al. (2005), we obtain a 2--10 keV flux of 
4.0$\times$10$^{-12}$ erg cm$^{-2}$ s$^{-1}$. This implies an 
X-ray/[O~{\sc iii}]$_{\rm 5007}$ ratio of $\sim$30, indicating that 
this source is well in the Compton-thin regime for Seyfert 2 galaxies 
(Bassani et al. 1999).

For IGR J19473+4452 we can calculate, after having taken into account the 
Galactic and intrinsic absorptions, the SFR and metallicity of this 
galaxy. Again following Kennicutt (1998), we determine a SFR of 
2.1$\pm$0.2 $M_\odot$ yr$^{-1}$ from the reddening-corrected H$_\alpha$ 
luminosity of (2.64$\pm$0.18)$\times$10$^{41}$ erg s$^{-1}$. The method 
(again in Kennicutt 1998) which instead uses the extinction-corrected [O 
{\sc ii}] luminosity, (5.3$\pm$1.3)$\times$10$^{42}$ erg s$^{-1}$, 
gives a SRF of 7$\pm$3 $M_\odot$ yr$^{-1}$, which is larger 
than, but still consistent with (at the 90\% confidence level) 
that derived using the H$_\alpha$ emission line flux.

Moreover, the detection of [O {\sc ii}], [O {\sc iii}] and H$_\beta$ also 
allows us to infer the gaseous oxygen abundance of this galaxy. In this 
occurrence also, the application of the Kobulnicky et al.'s (1999) method 
implies a basically solar oxygen abundance. Similar results are obtained 
using the [N~{\sc ii}]/H$_\alpha$ flux ratio method (Kewley \& Dopita 
2002).

\subsection{The nature of optically identified {\it INTEGRAL} sources}

Summing up all the knowledge available in the literature, at present 
(November
2005) 16 unknown or newly-discovered {\it INTEGRAL} sources were
identified by means of optical spectroscopy. These are 2 LMXBs (Paper I;
Roelofs et al. 2004), 7 HMXBs (this work; Reig et al. 2005; Halpern \&
Gotthelf 2004; Torrej\'on \& Negueruela 2004; Negueruela et al. 2005 and
references therein), 1 CV (Cieslinski et al. 1994; Masetti et al. 2005)
and 6 AGNs (this work; Paper I; Torres et al. 2004). In percentages, these
numbers translate into 56\% of XRBs (with 78\% of them being HMXBs and
22\% being LMXBs), 38\% of AGNs and 6\% of CVs.

If we compare these numbers with those coming out of the group of the 95 
identified objects belonging to the 1$^{\rm st}$ IBIS/{\it INTEGRAL} 
survey (Bird et al. 2004), that is, 80\% of XRBs (with only 30\% of them 
being HMXBs in this sample), 5\% of AGNs and 5\% of CVs, we see that a 
substantial fraction of the {\it INTEGRAL} sources identified {\it a 
posteriori} through opical spectroscopy and lying in the Galactic Plane is 
made of background AGNs. Albeit these small numbers do not allow us to 
perform an in-depth statistical analysis of the sample, we put forward the 
idea that {\it INTEGRAL} is a fundamental instrument with which to explore 
the Zone of Avoidance of the Galaxy not only for Galactic sources but 
also, and apparently mainly, for objects lying beyond the Galaxy. Besides, 
within the class of XRBs, a larger number of HMXBs, with respect to that 
of LMXBs, is detected among the unknown {\it INTEGRAL} sources.

Walter et al. (2004) already noted that {\it INTEGRAL} allowed the 
discovery of a new population of absorbed transient supergiant HMXBs; 
moreover, {\it INTEGRAL} doubled the number of known Galactic HMXBs with 
supergiant companion (Walter et al. 2005). We stress here that, 
equivalently, this hard X-ray telescope is also allowing us to pin down 
new AGNs lying in a strip of the sky which up to now has been poorly, or 
at least not carefully, explored in a systematic way.

\section{Conclusions}

In a sequel of the work started in Paper I, we have
identified four more {\it INTEGRAL} sources by means of optical
spectroscopy acquired at the Astronomical Observatory of Bologna in Loiano
(Italy).

We determined their nature as follows: (i) IGR J12391$-$1610 is an X-ray 
luminous Type 2 Seyfert galaxy in the Compton-thick regime, located at $z$ 
= 0.036, with solar metallicity and SFR $\sim$ 10 $M_\odot$ yr$^{-1}$; 
(ii) IGR J18406$-$0539 is a Be/X HMXB at a distance of $\sim$1.1 kpc from 
Earth and marginally consistent with being located
in the Sagittarius Arm of the Galactic Disk; (iii) 2E 1853.7+1534 is a 
luminous Type 1 Seyfert galaxy at $z$ = 0.084 with a central black hole 
of mass $\sim$1.4$\times$10$^{8}$ $M_\odot$; (iv) IGR J19473+4452 is 
a bright, Compton-thin regime, Seyfert 2 galaxy at $z$ = 0.053 with 
solar metallicity and SFR $\sim$ 2 $M_\odot$ yr$^{-1}$. 

We have also determined the nature and redshift ($z$ = 0.071) of a 
starburst/H {\sc ii} galaxy located 1$'$ east of LEDA 170194, the optical 
counterpart to IGR J12391$-$1610; we moreover regard as unlikely any 
contribution of that field galaxy to the total X-ray emission detected as 
IGR J12391$-$1610.

The statistical analysis of the (admittedly small, but growing) number of
{\it INTEGRAL} unknown or newly discovered sources, the nature of which is
being pinpointed through optical spectroscopy, shows that a substantial
fraction of them is of extragalactic origin. This underscores the
importance of hard X-ray observations for the study of background AGNs
located beyond the Zone of Avoidance of the Galaxy.

\begin{acknowledgements}

We are indebted with the anonymous referee for very useful comments and 
remarks which helped us to substantially improve the paper.
We thank A. De Blasi and R. Gualandi for the assistance in Loiano, and 
S. Marinoni, S. Bernabei, S. Galleti and I. Bruni for having performed 
part of the observations in Loiano as
a Service Mode run. This research has made use of the NASA Astrophysics
Data System Abstract Service, of the NASA/IPAC Extragalactic Database
(NED), and of the NASA/IPAC Infrared Science Archive, which are operated
by the Jet Propulsion Laboratory, California Institute of Technology,
under contract with the National Aeronautics and Space Administration.
This research has also made use of the SIMBAD database operated at CDS,
Strasbourg, France, and of the HyperLeda catalogue operated at the
Observatoire de Lyon, France. NM thanks the European Southern Observatory 
in Vitacura, Santiago (Chile) for the pleasant hospitality during the
preparation of this paper.

\end{acknowledgements}

\end{document}